\documentclass[useAMS,usenatbib]{mn2e}
\usepackage{times}
\usepackage{epsfig}
\usepackage[usenames]{color}

\title[Global dynamics of ADAFs with outflow]{Global dynamics of advection-dominated accretion flows with magnetically driven outflow}
\author[S. Li and X. Cao]{Shuang-Liang
Li\thanks{E-mail:lisl@shao.ac.cn}, Xinwu
Cao\thanks{E-mail:cxw@shao.ac.cn}\\
Key Laboratory for Research in Galaxies and Cosmology, Shanghai
Astronomical Observatory, Chinese Academy of Sciences,\\ 80 Nandan
RD, Shanghai, 200030, China}

\begin{document}

\date{}

\pagerange{\pageref{firstpage}--\pageref{lastpage}} \pubyear{2009}

\maketitle

\label{firstpage}

\begin{abstract}

We study the global dynamics of advection-dominated accretion
flows (ADAFs) with magnetically driven outflows. A fraction of
gases in the accretion flow is accelerated into the outflows,
which leads to decreasing of the mass accretion rate in the
accretion flow towards the black hole. We find that the
$r$-dependent mass accretion rate is close to a power-law one,
$\dot{m} {\propto} r^s$, as assumed in the advection-dominated
inflow-outflow solution (ADIOS), in the outer region of the ADAF,
while it deviates significantly from the power-law $r$-dependent
accretion rate in the inner region of the ADAF. It is found that
the structure of the ADAF is significantly changed in the presence
of the outflows. The temperatures of the ions and electrons in the
ADAF decreases in the presence of outflows, as a fraction of
gravitational power released in the ADAF is tapped to accelerate
the outflows.

\end{abstract}


\section{Introduction\label{intro}}

It is widely believed that many astrophysical objects are powered
by mass accretion on to black holes. The standard geometrically
thin, optically thick accretion disc model can successfully
explain most of the observational features in active galactic
nuclei (AGN) and X-ray binaries \citep{s1973}.  In the standard
thin model, the motion of the matter in the accretion disc is
nearly Keplerian, and the gravitational energy released in the
disc is radiated away locally. An alternative accretion disc
model, namely, the advection-dominated accretion flow (ADAF)
model, was suggested for the black holes accreting at very low
rates \citep{1977ApJ...214..840I,n1994}. In the ADAF model, only a
small fraction of the gravitational energy released in the
accretion flow is radiated away due to inefficient cooling, and
most of the energy is stored in the accretion flow and advected to
the black hole. The ADAFs are optically thin and hot (comparable
with the virial temperature of the gases in the flows), which
radiate mostly in X-ray waveband \citep*[see][for a review and
references therein]{2008NewAR..51..733N}. This model can
successfully explain the main observational features of black hole
X-ray binaries and low-luminosity AGN (LLAGN)
\citep*[e.g.,][]{n1994,n1995a,g1999,q1999,y2003,h2008}. As the
Bernoulli parameter of an ADAF is positive, the ADAF is likely to
have an outflow, which was confirmed by numerical simulations and
also supported by observations
\citep{s2001,i2003,m2006,g1999,q1999,y2003}.

\citet{b1999} proposed a self-similar advection-dominated
inflow-outflow solution (ADIOS) for the ADAF with winds. In ADIOS
model, the mass accretion rate is no longer a constant and is
assumed to be a power-law dependence of radius
($\dot{m}\propto{r^{s}}, 0<s<1$), which is an important ingredient
in most of the follow-up works \citep*[e.g.,][]{q1999b,y2003,x2005}.
Motivated by the results of numerical simulations on accretion
discs, \citet{x2008} investigated the influence of outflows on the
accretion flow based on a 1.5-dimensional description of the
accretion flow. They suggested that their solutions can be described
by a power-law $r$-dependent mass accretion rate fairly well.

Magnetic fields are believed in accretion flows, and the
magnetorotational instability (MRI) provides the source of viscosity
in accretion flows \citep{b1991,b1998}. The outflows/jets can be
driven by the large-scale ordered magnetic fields threading the
accretion disc \citep{b1982}. The physics of magnetically
accelerated outflows has been extensively explored in many previous
works
\citep*[e.g.,][]{1994A&A...287...80C,c2002a,1995ApJ...452L..41K,
1999ApJ...522..727K,2002ApJ...565.1035K,1998ApJ...499..329O,2001ApJ...553..158O,2007MNRAS.375..548N,2007MNRAS.375..513M,2007MNRAS.375..531M}.
Such outflows/jets magnetically driven from the accretion discs
provide an efficient angular momentum loss mechanism for accretion
discs \citep*[see][for a review and references
therein]{2008arXiv0804.3096S}. The structure of a standard thin
disc/ADAF may be altered by the magnetically driven outflows
\citep*[e.g.,][]{2008A&A...482....1L,li09}. In this work, we
investigate the global structure of an ADAF with magnetically driven
outflows/jets.

\section{Model}\label{equations}

We consider a steady ADAF with magnetically driven outflows/jets
surrounding a black hole in this work.

The continuity equation is
\begin{equation}
\frac{d}{dR}(2{\pi}R\Sigma v_{\rm R})+4\pi R\dot{m}_{\rm w}=0,
\label{mass}
\end{equation}
where $v_{\rm R}$ is the radial velocity, $\Sigma=2H\rho$ is the
surface density of the accretion flow, and $\dot{m}_{\rm w}$ is the
mass loss rate from unit surface area of accretion flow. The
half-thickness of the disc $H$ is given by $H=c_{\rm s}/\Omega_{\rm
K}$, and $\Omega_{\rm K}$ is the Keplerian angular velocity. The
sound speed $c_{\rm s}=(P/\rho)^{1/2}$, and the total pressure $P$
is the sum of the gas pressure and the magnetic pressure: $P=P_{\rm
gas}+P_{\rm m}=P_{\rm i}+P_{\rm e}+P_{\rm m}$ ($P_{\rm i}$ and
$P_{\rm e}$ are the ion pressure and the electron pressure
respectively).

In this work, we adopt the Paczy\'{n}ski--Wiita potential
\begin{equation}
\psi=-\frac{GM}{R-R_{\rm g}}
\end{equation}
to simulate the general relativistic effects of a Schwarzschild
black hole, where $M$ is the mass of the black hole, and $R_{\rm
g}=2GM/c^2$ is the gravitational radius \citep{p1980}.

The radial momentum equation is
\begin{equation}
v_{\rm R} \frac{dv_{\rm R}}{dR}-R(\Omega^2-\Omega_{\rm
K}^2)+\frac{1}{\rho}\frac{dP}{dR}-g_{\rm m}=0, \label{radial}
\end{equation}
where $\Omega$ is the angular velocity of the accretion flow. The
radial magnetic force is given by
\begin{equation}
g_{\rm m}=\frac{{B_{\rm r}^{\rm s}}B_{\rm z}}{2\pi\Sigma},
\label{gm}
\end{equation}
where $B_{r}^{\rm s}$ and $B_{z}$ are the radial and vertical
components of the magnetic fields at the disc surface.

The angular momentum equation reads
\begin{equation}
v_{{\rm}R}\frac{d({\Omega}R^2)}{dR}-\frac{1}{{\rho}HR}\frac{d}{dR}(R^{2}H\tau_{r\varphi})+\frac{T_{\rm
m}}{\Sigma}=0, \label{angular}
\end{equation}
where $\alpha$-viscosity $\tau_{r\varphi}=-{\alpha}P$ is adopted
\citep{s1973}, and $T_{\rm m}$ is the magnetic torque exerted on the
accretion flow due to the outflows/jets. The outflow is accelerated
by the magnetic fields threading the rotating accretion disc, and
therefore the torque $T_{\rm m}$ can be calculated with
\begin{equation}
T_{\rm m}=2\dot{m}_{\rm w}\Omega(R_{\rm d})(R_{\rm A}^2-R_{\rm
d}^2), \label{tm}
\end{equation}
where $\dot{m}_{\rm w}$ is the mass loss rate due to the outflow,
$R_{\rm d}$ is the radius of the footpoint of the field line, and
$R_{\rm A}$ is the Alfv\'{e}n point \citep*[see, e.g.,][for the
details]{c2002a}.

The energy equations for ions and electrons are given by
\begin{equation}
{\rho}v_{{\rm}R}(\frac{d\varepsilon_{\rm e}}{dR}-\frac{P_{\rm
e}}{\rho^2}\frac{d\rho}{dR})-{\delta}q^{+}-q_{\rm
ie}+q^{-}+\frac{2\dot{m}_{\rm w}\varepsilon_{\rm e}}{2H}=0,
\label{energy1}
\end{equation}
and
\begin{equation}
{\rho}v_{\rm R}(\frac{d\varepsilon_{\rm i}}{dR}-\frac{P_{\rm
i}}{\rho^2}\frac{d\rho}{dR})-{(1-\delta)}q^{+}+q_{\rm
ie}+\frac{2\dot{m}_{\rm w}\varepsilon_{\rm i}}{2H}=0,
\label{energy2}
\end{equation}
respectively, where the parameter $\delta$ describes the fraction of
the viscously dissipated energy that goes directly into electrons in
the accretion flow, and the specific internal energy of electrons
and ions are given by
\begin{equation}
\varepsilon_{\rm e}=\frac{1}{\gamma_{\rm e}-1}\frac{kT_{\rm
e}}{\mu_{\rm e}{m_{\rm H}}},
\end{equation}
\begin{equation} \varepsilon_{\rm i}=\frac{1}{\gamma_{\rm
i}-1}\frac{kT_{\rm i}}{\mu_{\rm i}{m_{\rm H}}},
\end{equation}
where $T_{\rm e}$ and $T_{\rm i}$ are the temperature of electrons
and ions respectively, and the mean molecular weight of the ions
and the electrons: $\mu_{\rm i}=1.23$, $\mu_{\rm e}=1.14$ are
adopted. The adiabatic indices of the electrons and ions,
$\gamma_{\rm e}$ and $\gamma_{\rm i}$, are given by
\begin{equation}
\gamma_{\rm e}=1+\theta_{\rm e}\left[\frac{3K_{3}(1/\theta_{\rm
e})+K_{1}(1/\theta_{\rm e})}{4K_{2}(1/\theta_{\rm
e})}-1\right]^{-1},
\end{equation}
\begin{equation}
\gamma_{\rm i}=1+\theta_{\rm i}\left[\frac{3K_{3}(1/\theta_{\rm
i})+K_{1}(1/\theta_{\rm i})}{4K_{2}(1/\theta_{\rm
i})}-1\right]^{-1},
\end{equation}
where $K^{,}s$ are the modified Bessel functions, and the
dimensionless electron and ion temperature are defined as:
$\theta_{\rm e}=kT_{\rm e}/(m_{\rm e}c^2)$ and $\theta_{\rm
i}=kT_{\rm i}/(m_{\rm p}c^2)$ \citep{n1995b}. The energy dissipation
rate per unit volume is given by
$q^{+}=-{\alpha}PR{d{\Omega}}/{dR}$, and $q_{\rm ie}$ indicates the
energy transfer rate from ions to electrons through Coulomb
collisions, which is given by \citep{s1983}
\begin{displaymath}
q_{\rm ie}=\frac{3}{2}\frac{m_{\rm e}}{m_{\rm p}}n_{\rm e}n_{\rm
i}\sigma_{\rm T}c\frac{(kT_{\rm i}-kT_{\rm
e})}{K_{2}(1/\theta_{\rm e})K_{2}(1/\theta_{\rm i})}\rm{ln}\Lambda
\end{displaymath}

\begin{equation}
\times\left[\frac{2(\theta_{\rm e}+\theta_{\rm
i})^2+1}{(\theta_{\rm e}+\theta_{\rm
i})}K_{1}\left(\frac{\theta_{\rm e}+\theta_{\rm i}}{\theta_{\rm
e}\theta_{\rm i}}\right)+2K_{0}\left(\frac{\theta_{\rm
e}+\theta_{\rm i}}{\theta_{\rm e}\theta_{\rm i}}\right)\right],
\end{equation}
where the Coulomb logarithm $\rm{ln}\Lambda=20$, and $q^{-}$ is the
radiative cooling rate consisting of synchrotron, bremsstrahlung,
and Compton cooling \citep*[see,][for details]{n1995b,m2000}.

The dynamical properties of magnetically driven outflows/jets from
an accretion disk can be investigated by solving a set of
magneto-hydrodynamical (MHD) equations if the magnetic field
configuration of the disk and suitable boundary conditions at the
disk surface are supplied
\citep*[e.g.,][]{1994A&A...287...80C,c2002a,1995ApJ...452L..41K,1999ApJ...522..727K,2002ApJ...565.1035K,1998ApJ...499..329O,2001ApJ...553..158O}.
However, the generation and maintenance of large-scale magnetic
fields of the disk is still quite unclear
\citep*[e.g.,][]{1994MNRAS.267..235L,1994MNRAS.268.1010L,1996MNRAS.281..219T}.
In this work, we focus on how the dynamics of the ADAF is affected
by the presence of magnetically driven outflows/jets. For
simplicity, we assume that large-scale magnetic field lines thread
the accretion disk, and the strength of the magnetic fields far from
the disc surface along the field line to be roughly self-similar:
\begin{equation}
B_{\rm p}(R)\sim B_{\rm pd}\left({\frac {R}{R_{\rm
d}}}\right)^{-\zeta}, \label{b_p}
\end{equation}
where $R_{\rm d}$ is the radius of the field footpoint at the disc
surface, $B_{\rm pd}$ is the strength of the poloidal component of
the field at the disc surface, $B_{\rm p}(R)$ is the field strength
at $R$ along the field line, the self-similar index $\zeta\ge 1$
\citep{b1982}, and $\zeta=4$ is adopted in the calculations of
\citet{1994MNRAS.268.1010L}.

The self-similar wind solution derived by
\citet{1994MNRAS.268.1010L} is only valid for slowly moving
(non-relativistic) outflows, which was extended for relativistic
jets by \citet{c2002a}. In this work, we model the magnetically
driven outflows/jets with the approach adopted in \citet{c2002a}. We
summarize the model as follows \citep*[see][for the
details]{c2002a}.

For a relativistic jet accelerated by the magnetic field of the
disc, the Alfv\'{e}n velocity is \citep{1969ApJ...158..727M,c1986}
\begin{equation}
v_{\rm A}=\frac {B_{\rm p}^{\rm A}}{{(4\pi\rho_{\rm A}\gamma_{\rm
j})}^{1/2}}, \label{v_A}
\end{equation}
where $B_{\rm p}^{\rm A}$ and $\rho_{\rm A}$ are the poloidal field
strength and the density of the outflow/jet at Alfv\'{e}n point, and
$\gamma_{\rm j}$ is the Lorentz factor of the bulk motion of the
outflows/jets. In this work, all our calculations of the
outflows/jets are in the special relativistic frame. The Alfv\'{e}n
velocity $v_{\rm A}\sim R_{\rm A}\Omega(R_{\rm d})$, where $R_{\rm
A}$ is the radius of the Alfv\'{e}n point along the field line, and
$\Omega(R_{\rm d})$ is the angular velocity of the accretion flow at
the field footpoint $R_{\rm d}$.

The mass and magnetic flux conservation along the field line
requires
\begin{equation}
{\frac {\dot{m}_{\rm w}}{B_{\rm pd}}}\simeq {\frac {\rho_{\rm
A}v_{\rm A}}{B_{\rm p}^{\rm A}}}, \label{m_b_cons}
\end{equation}
where $\dot{m}_{\rm w}$ is the mass loss rate in the outflow/jet
from unit surface area of the disc.

The final bulk velocity of the magnetically driven outflow/jet is
$\sim v_{\rm A}$, so the Lorentz factor of the outflow/jet is
\begin{equation}
\gamma_{\rm j}\simeq \left[1-\left({\frac {v_{\rm A}}{c}}\right)^2
\right]^{-{1\over 2}}. \label{gam_j}
\end{equation}

Combining equations (\ref{b_p})--(\ref{gam_j}), the mass loss rate
in the outflow/jet from the unit surface area of the disc is
\begin{equation}
\dot{m}_{\rm w}=\frac{B_{\rm pd}^2}{4{\pi}c}\left[\frac{R_{\rm
d}\Omega(R_{\rm d})}{c}\right]^{\zeta}\frac{\gamma_{\rm
j}^{\zeta}}{(\gamma_{\rm j}^2-1)^{(1+\zeta)/2}}. \label{mw}
\end{equation}

The origin of the ordered magnetic fields threading the disc is
still unclear. It was suggested that the magnetic fields can be
generated through dynamo processes in the disc
\citep*[e.g.,][]{s1973,1996MNRAS.281..219T,1998ApJ...501L.189A,1998ApJ...500..703R},
or the large-scale external magnetic fields are transported inward
by the accretion flow
\citep*[e.g.,][]{1976Ap&SS..42..401B,1994MNRAS.267..235L,2005ApJ...629..960S}.
For simplicity, we assume the strength of the large-scale magnetic
fields threading the disc to be comparable with that of the fields
in the accretion disc, $B_{\rm pd}\simeq B$. The magnetic pressure
is conventionally assumed to be proportional to the gas pressure in
the accretion flow. Thus, we have
\begin{equation}
P_{\rm m}=\frac{B^2}{8\pi}=\frac{1-\beta}{\beta}P_{\rm gas},
\label{pmag}
\end{equation}
where $\beta$ is the ratio of the gas pressure to the total
pressure, $B$ is the strength of the magnetic fields in the
accretion flow.

The magnetically driven outflow is described by the terminal
velocity of the outflow (Alfv\'{e}n velocity $v_{\rm A}$) when the
values of two parameters $\zeta$ and $\beta$ are specified.

\section{results}\label{results}

\begin{figure*}
\includegraphics[width=7.5cm]{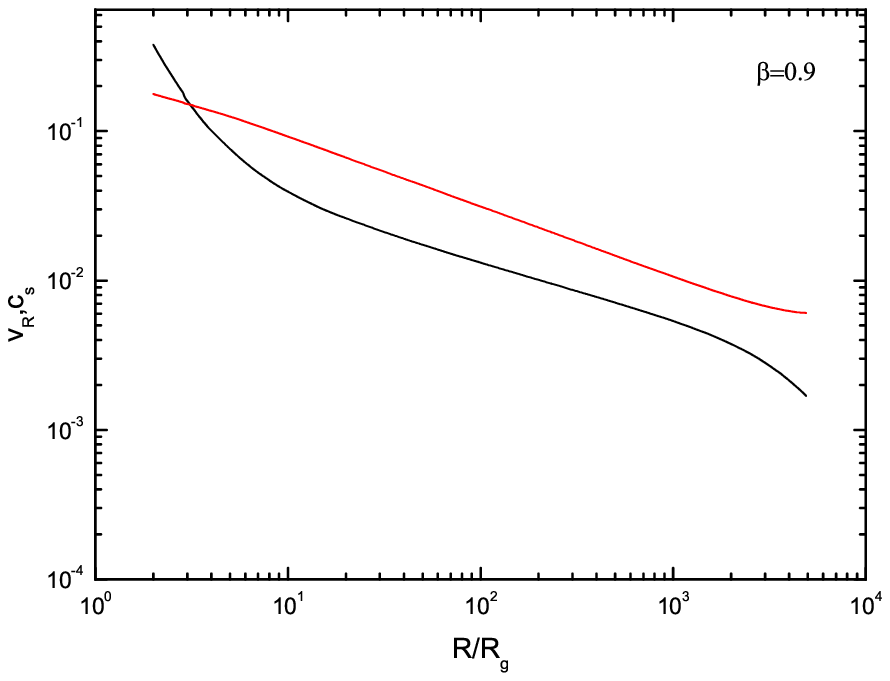}
\includegraphics[width=7.5cm]{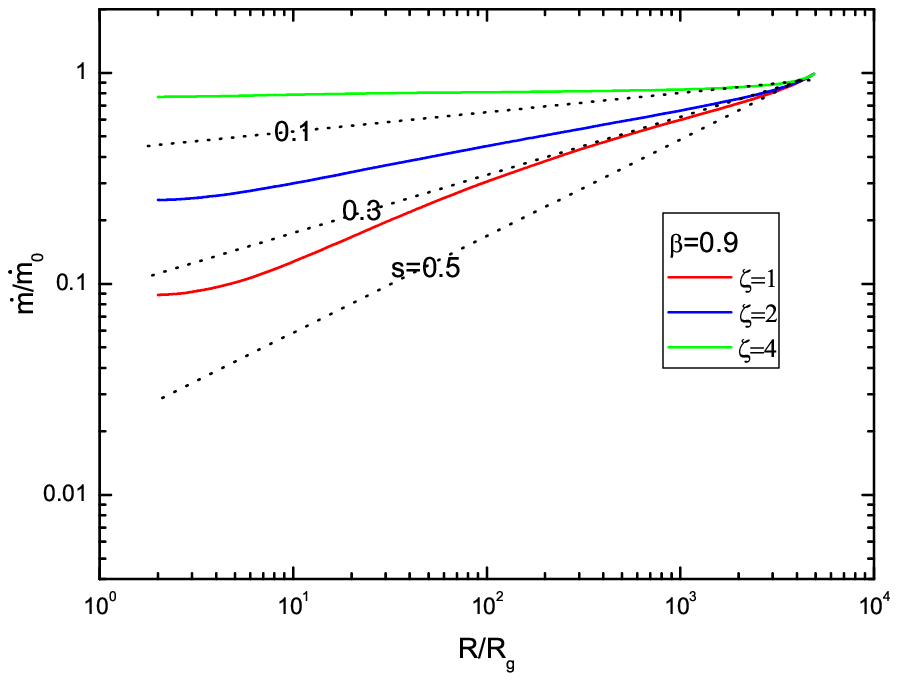}
\includegraphics[width=7.5cm]{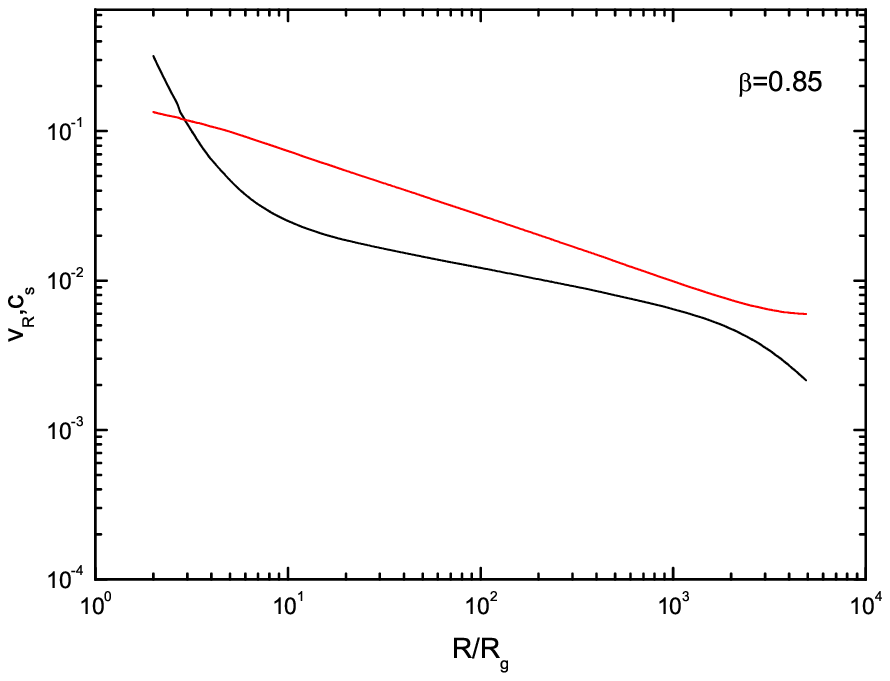}
\includegraphics[width=7.5cm]{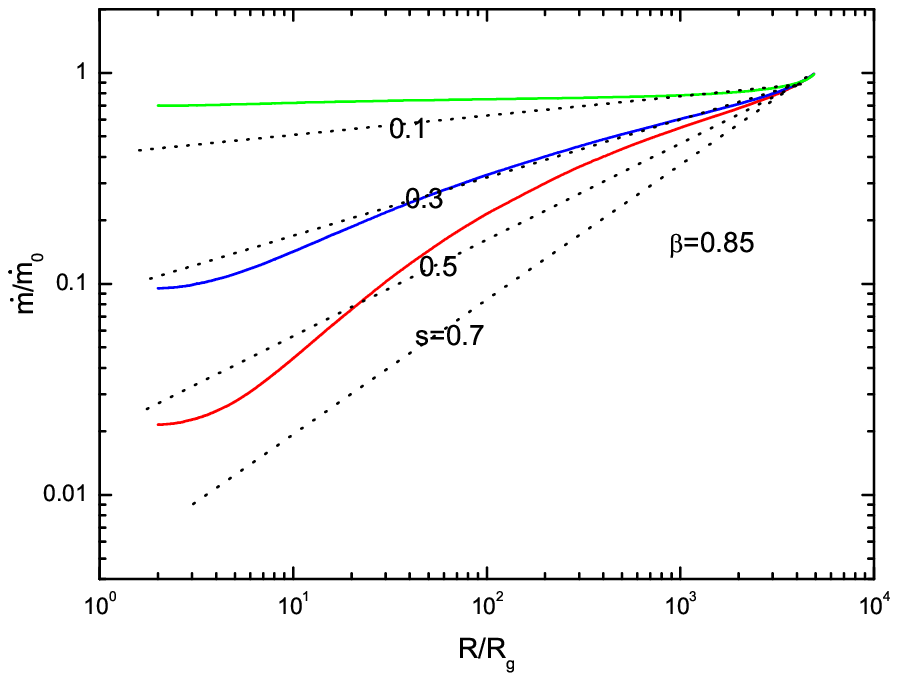}
\includegraphics[width=7.5cm]{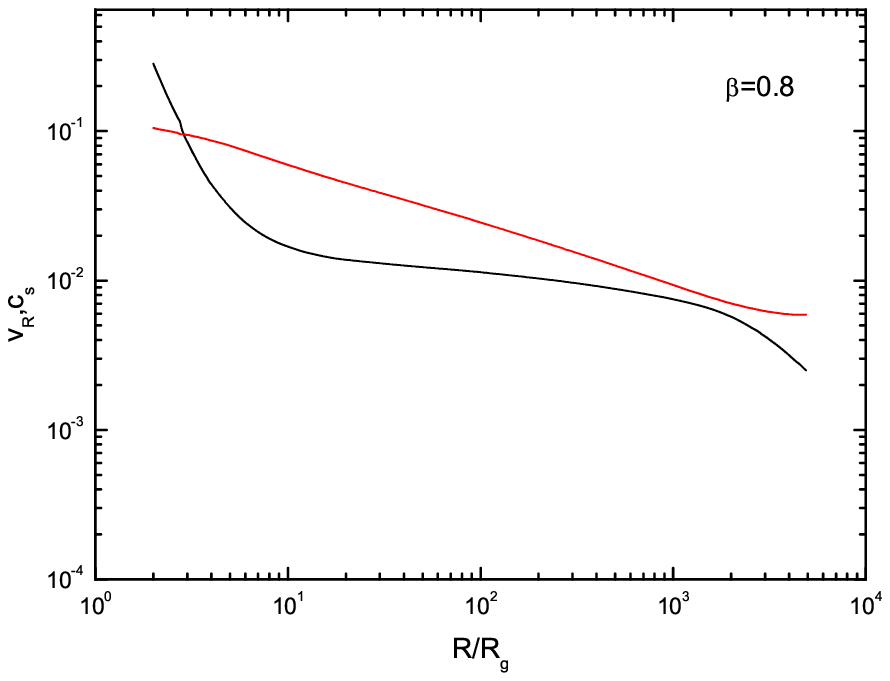}
\includegraphics[width=7.5cm]{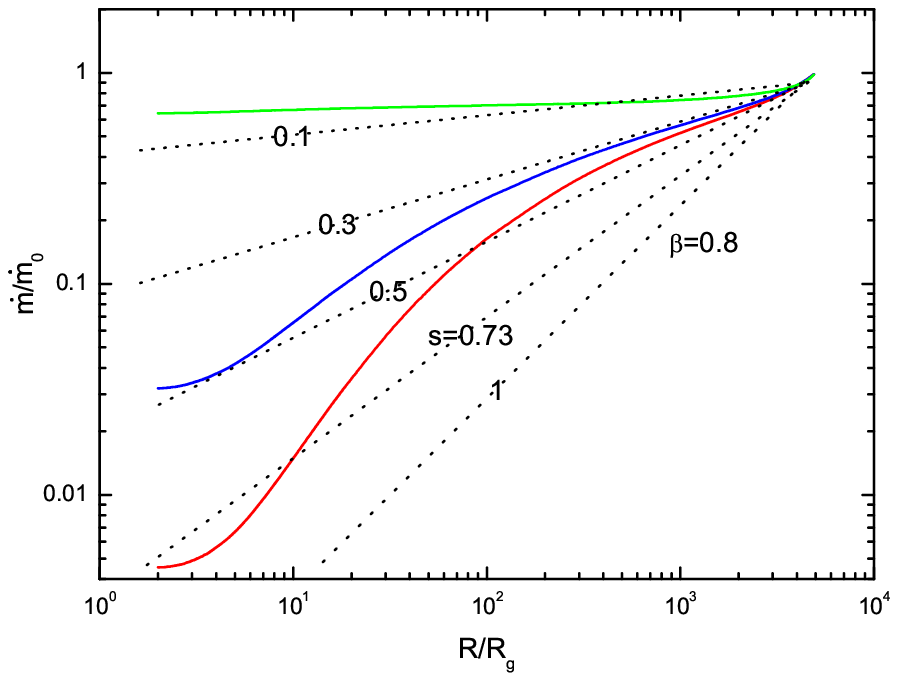}
\caption{The global solutions of ADAFs with magnetically driven
outflows with different parameters. The black and red lines in the
left figures are for the radial velocity $v_{\rm R}$ and the sound
speed $c_{\rm s}$ respectively. The dotted lines represent
power-law $r$-dependent mass accretion rates with different values
of $s$. The accretion rate at the outer radius ($r_{\rm
out}=5000$) is: $\dot{m}_0=10^{-5}$. \label{indexs1}}
\end{figure*}

\begin{figure*}
\includegraphics[width=7.5cm]{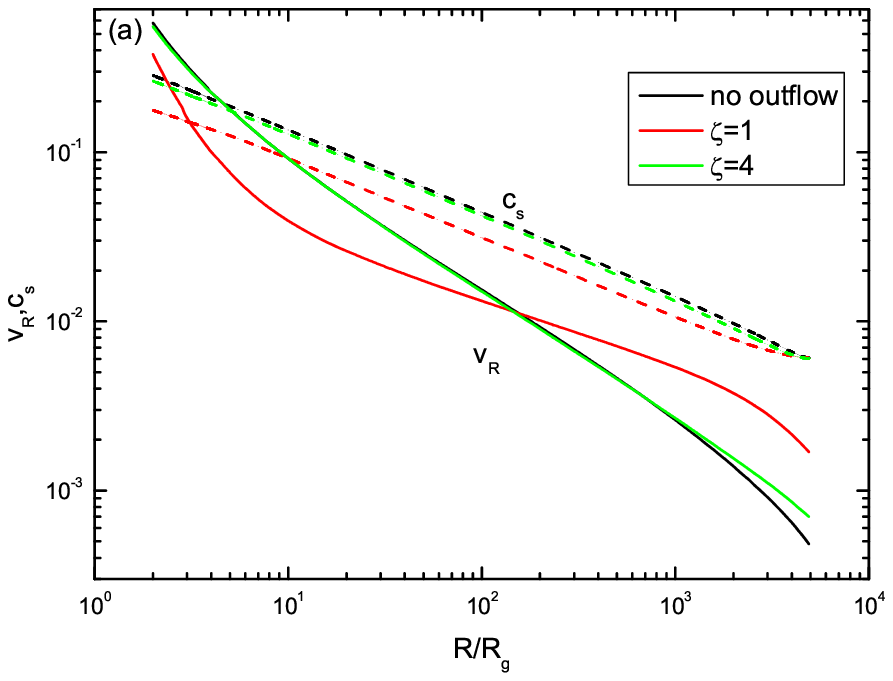}
\includegraphics[width=7.5cm]{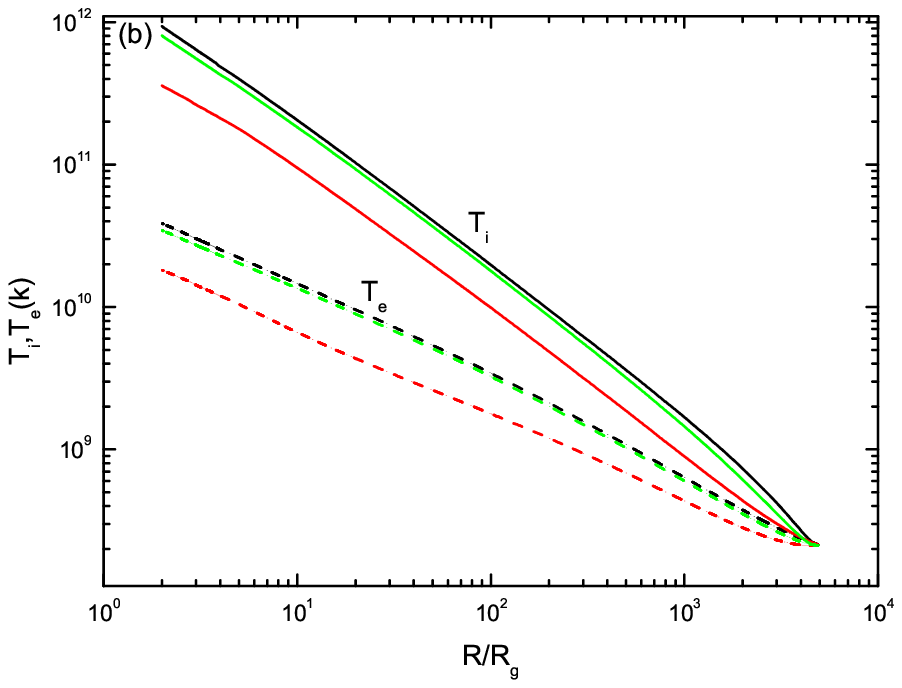}
\includegraphics[width=7.5cm]{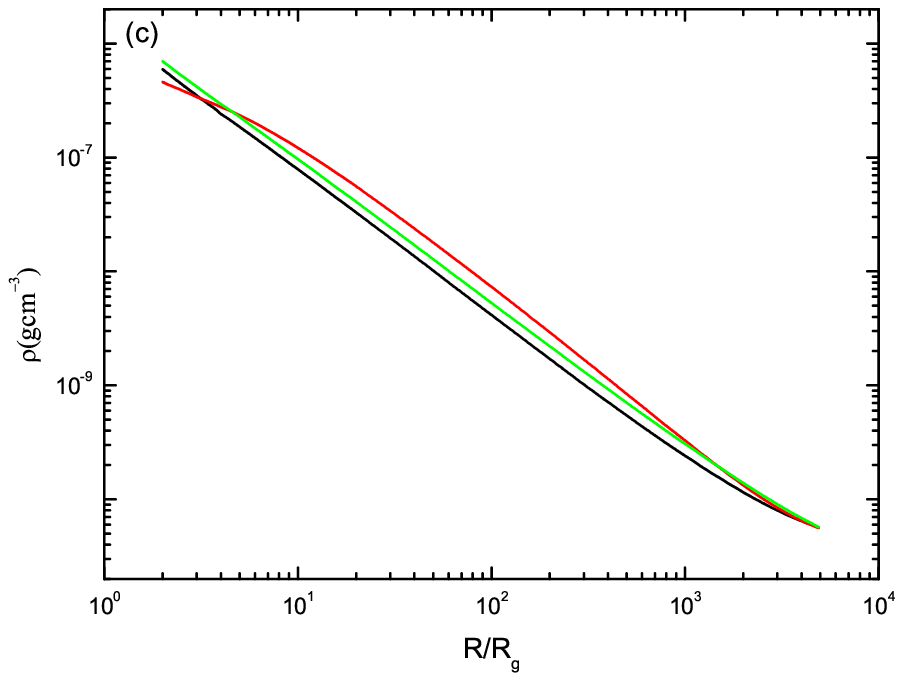}
\includegraphics[width=7.5cm]{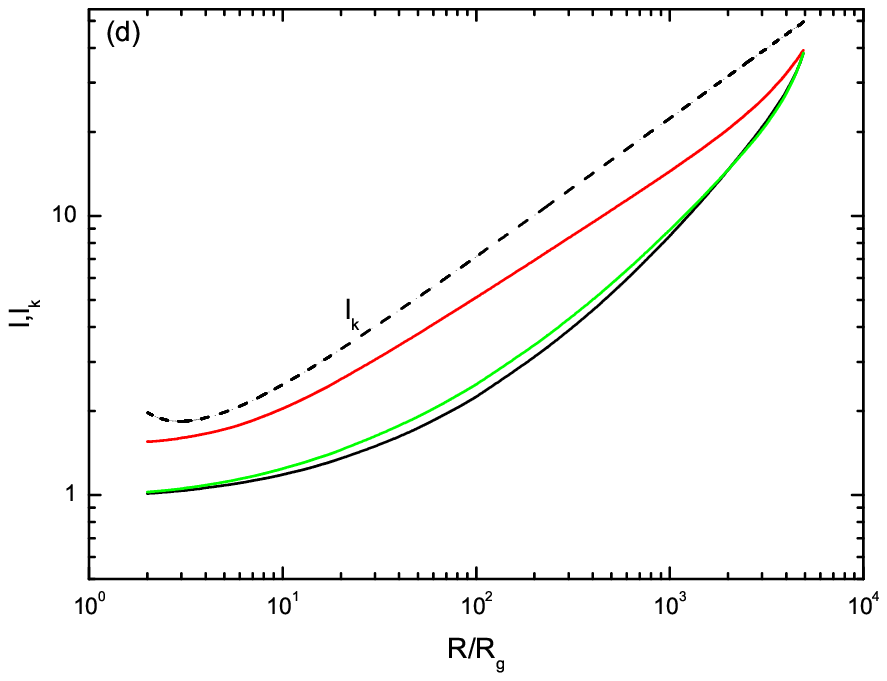}
\caption{The global structures of ADAFs with magnetically driven
outflows. We also plot the structures of ADAFs without outflows for
comparison (black lines).
 \label{zeta1}}
\end{figure*}

\begin{figure*}
\includegraphics[width=7.5cm]{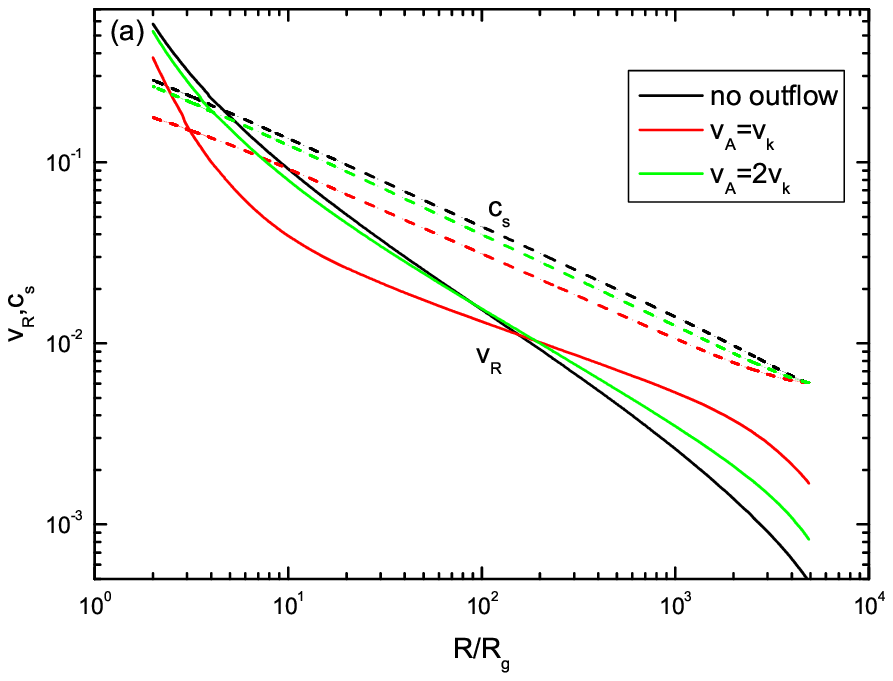}
\includegraphics[width=7.5cm]{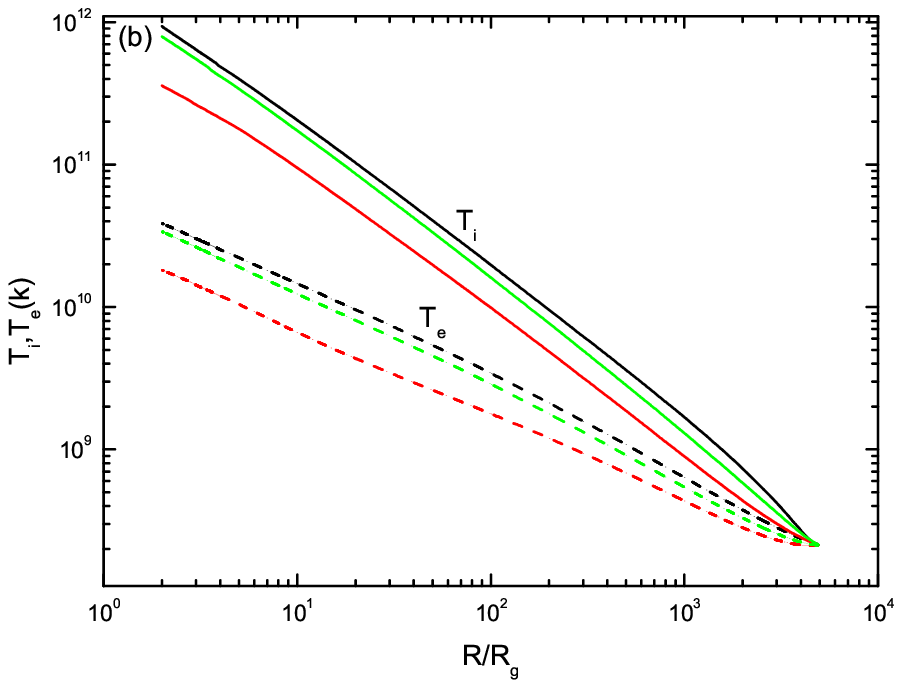}
\includegraphics[width=7.5cm]{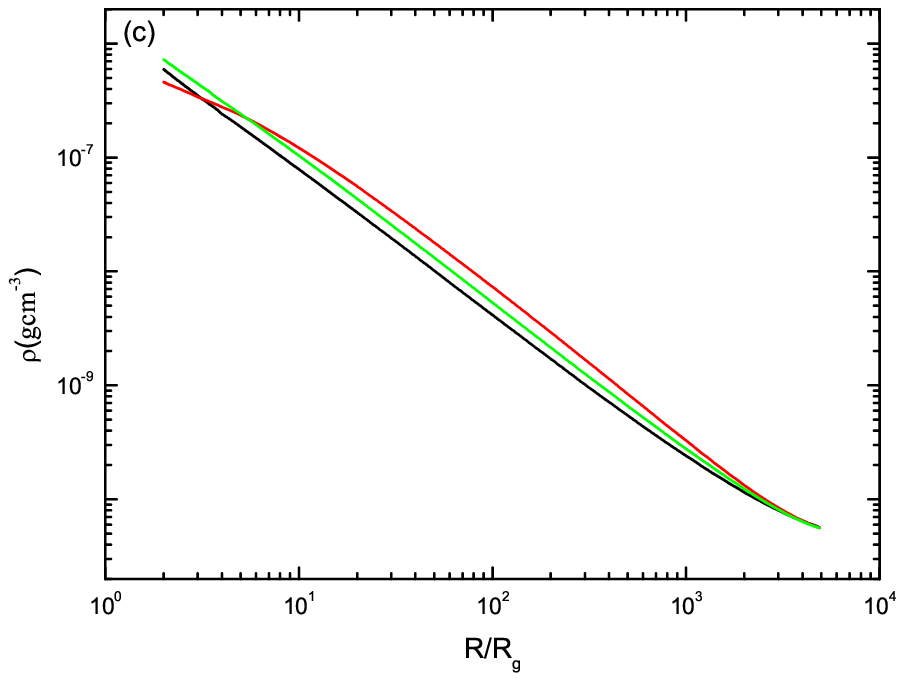}
\includegraphics[width=7.5cm]{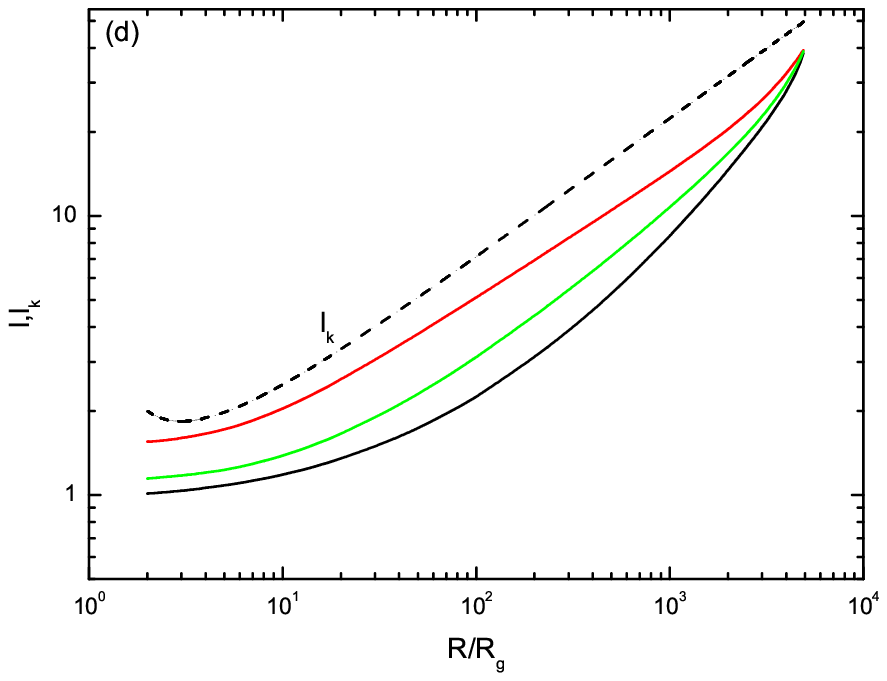}
\caption{The comparison of global structures of ADAFs with
different terminal velocities of the outflows. We also plot the
structures of ADAFs without outflows for comparison (black lines).
\label{va1}}
\end{figure*}

\begin{figure*}
\includegraphics[width=7.5cm]{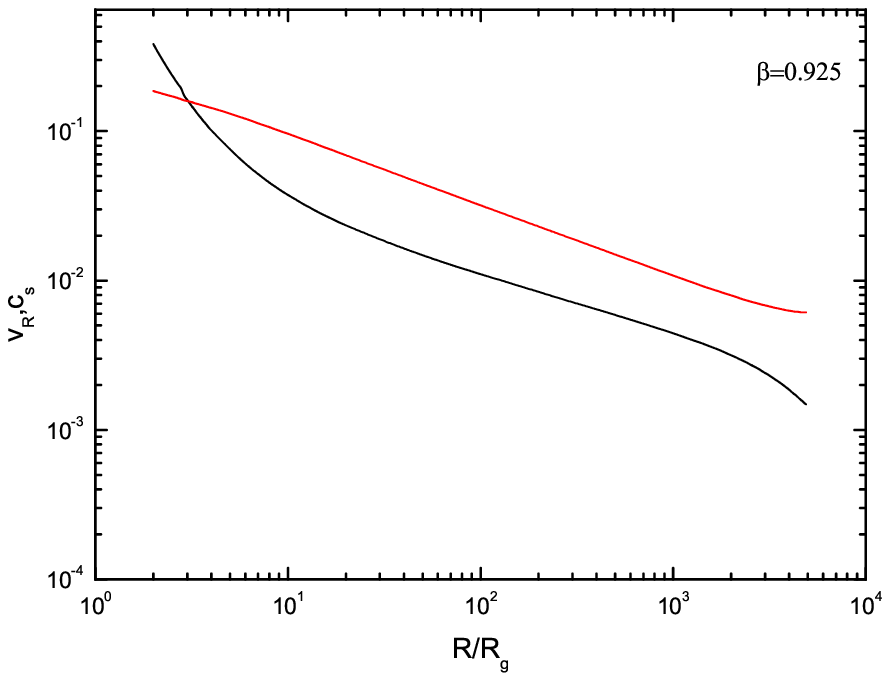}
\includegraphics[width=7.5cm]{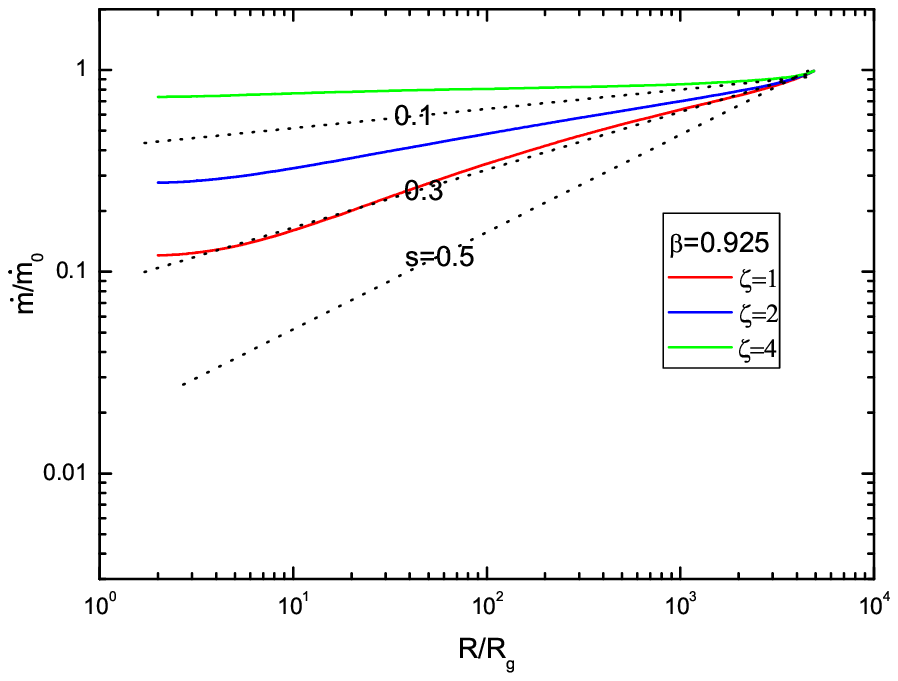}
\includegraphics[width=7.5cm]{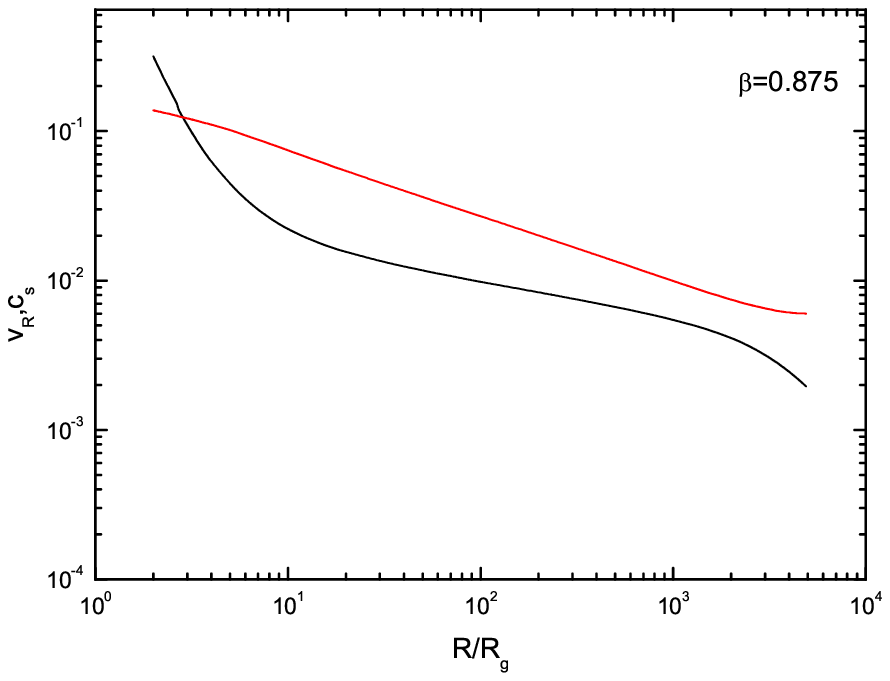}
\includegraphics[width=7.5cm]{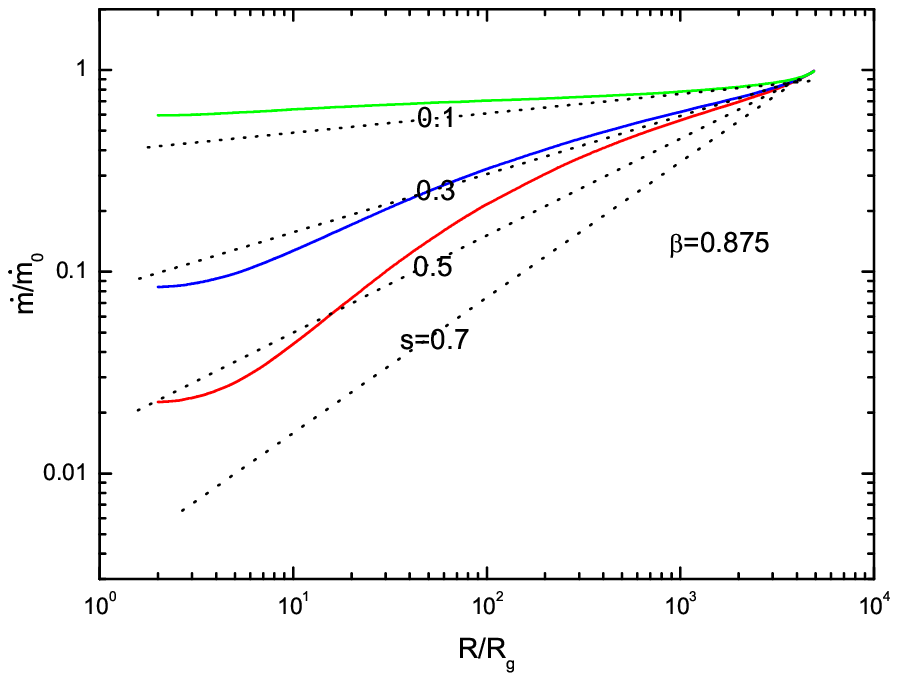}
\includegraphics[width=7.5cm]{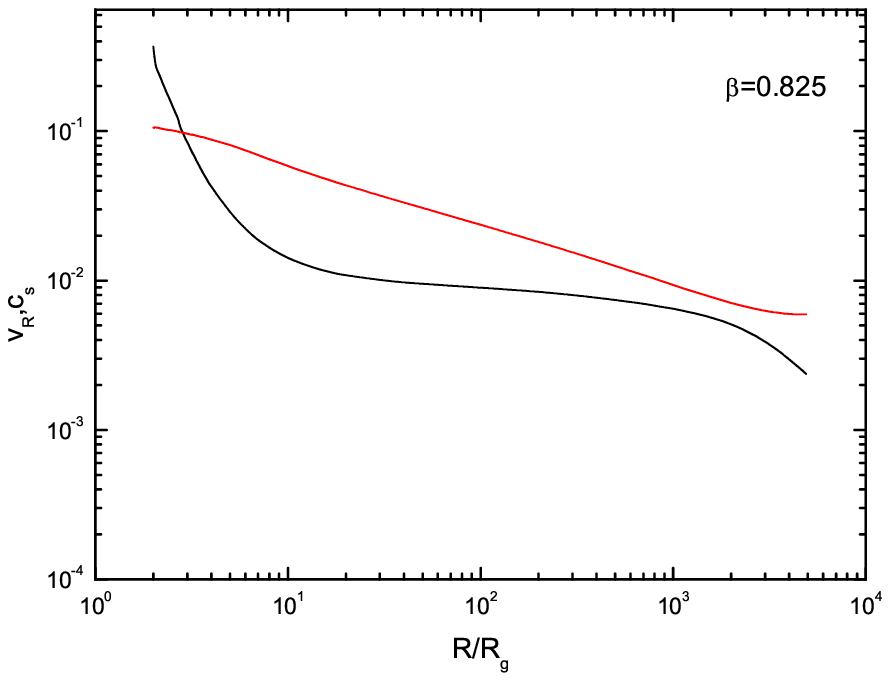}
\includegraphics[width=7.5cm]{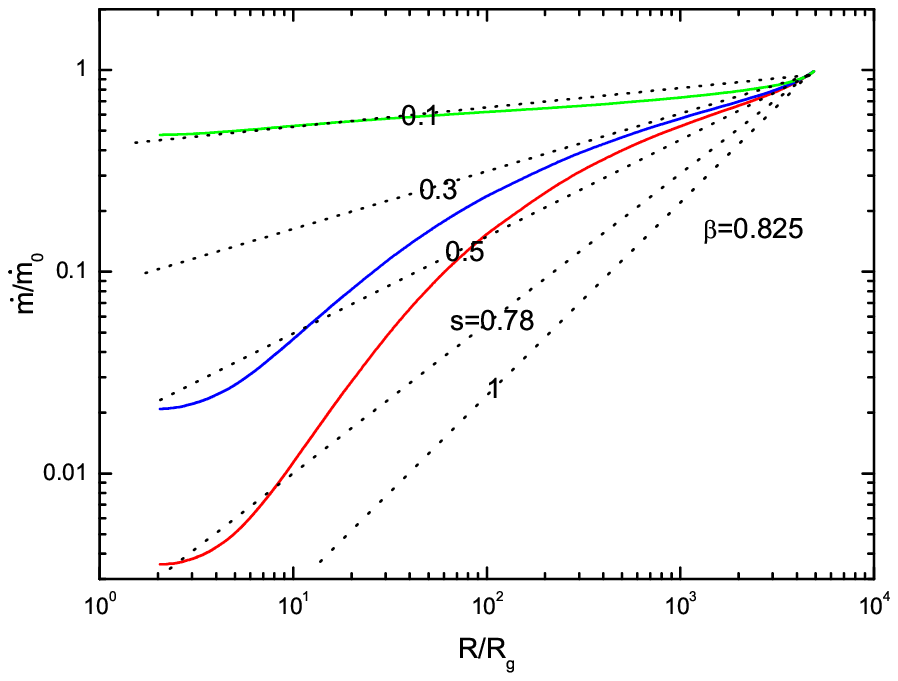}
\caption{The same as Fig. \ref{indexs1}, but the accretion rate at
the outer radius, $\dot{m}_0=10^{-2}$, is adopted in the
calculations. \label{indexs2}}
\end{figure*}

\begin{figure*}
\includegraphics[width=7.5cm]{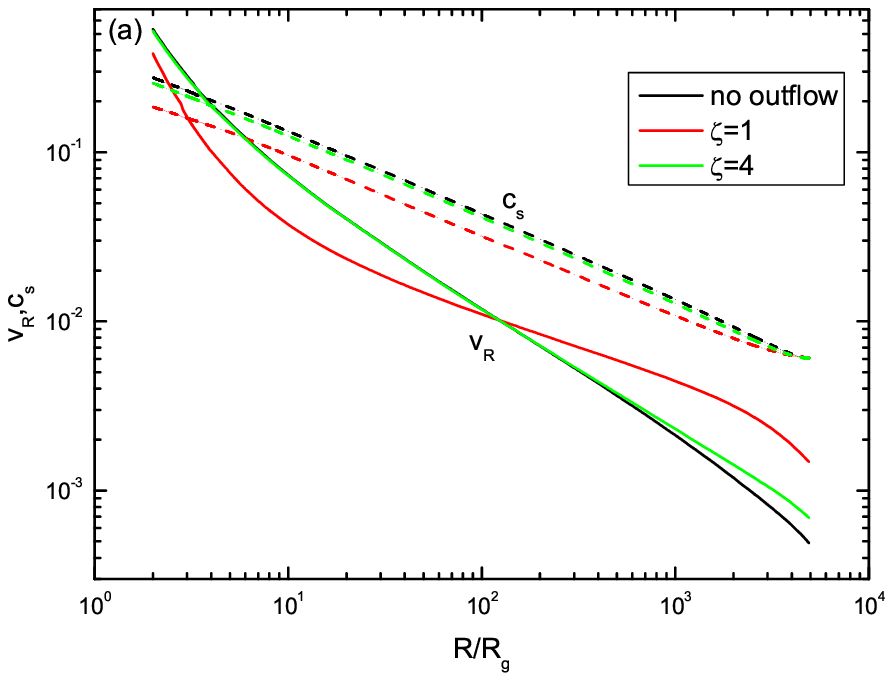}
\includegraphics[width=7.5cm]{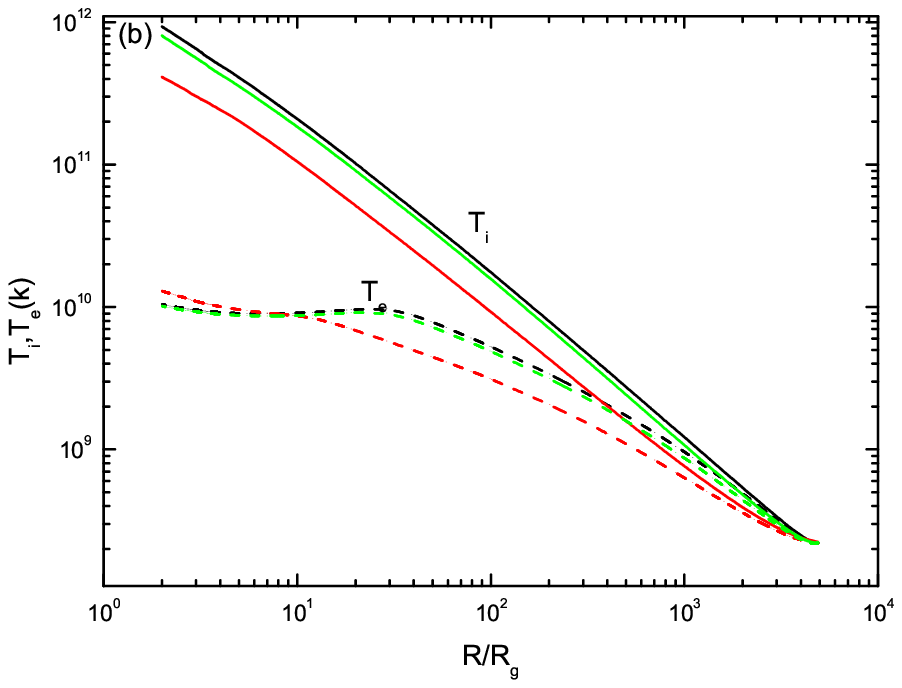}
\includegraphics[width=7.5cm]{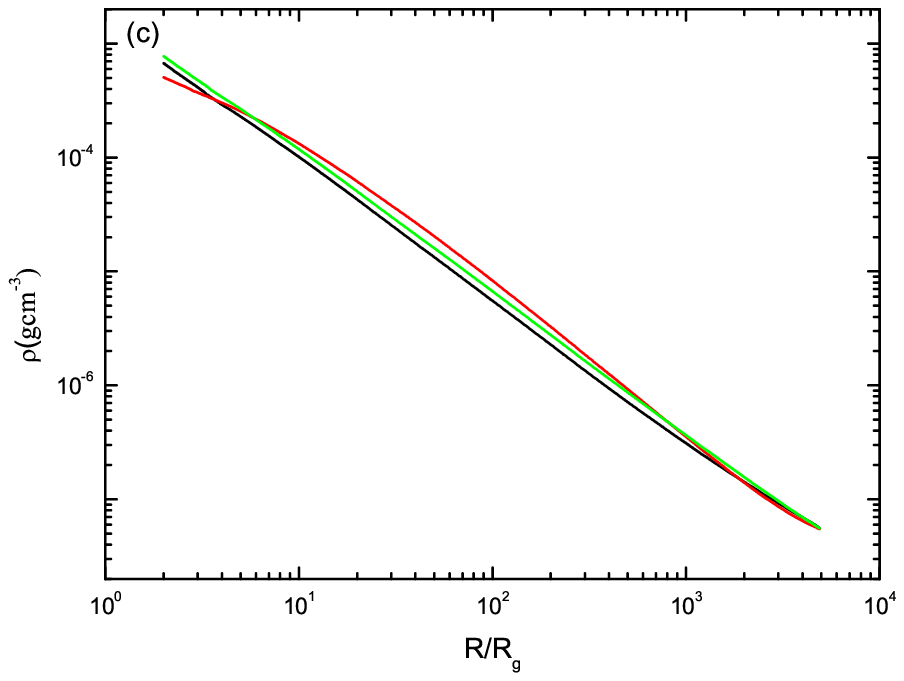}
\includegraphics[width=7.5cm]{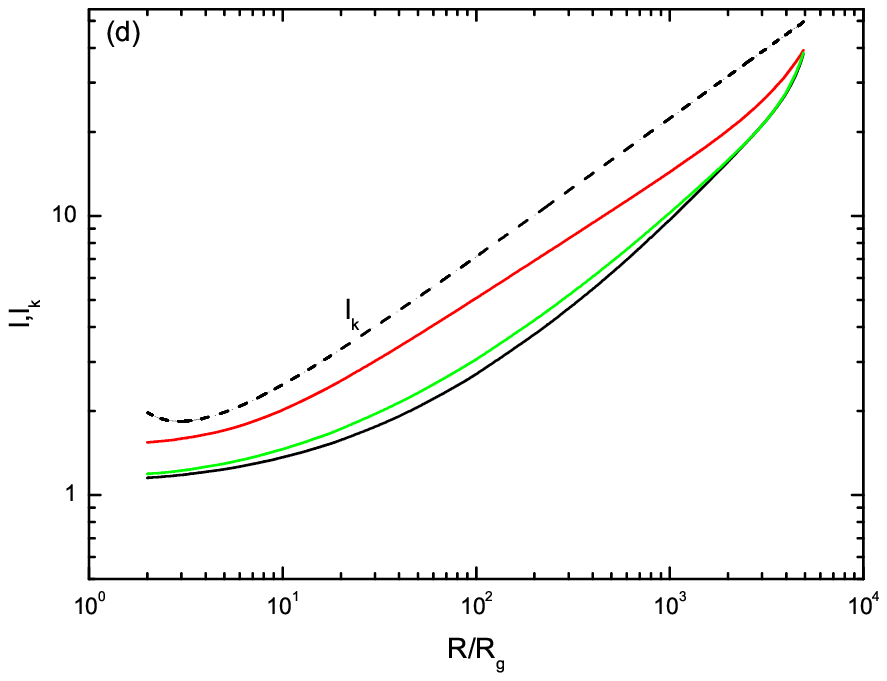}
\caption{The same as Fig. \ref{zeta1}, but the accretion rate at the
outer radius $\dot{m}_0=10^{-2}$, and $\beta=0.925$, are adopted.
\label{zeta2}}
\end{figure*}

\begin{figure*}
\includegraphics[width=7.5cm]{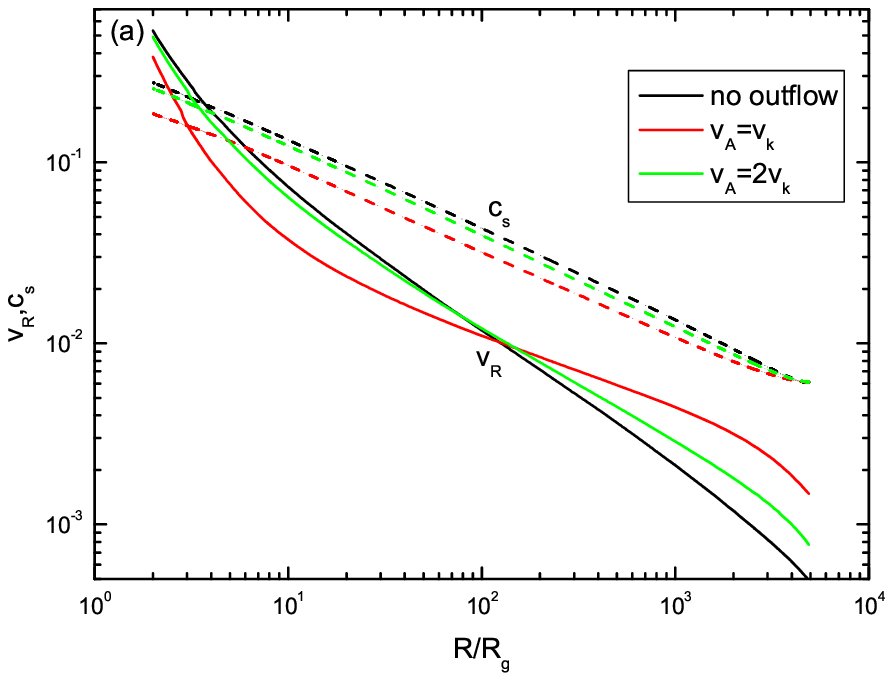}
\includegraphics[width=7.5cm]{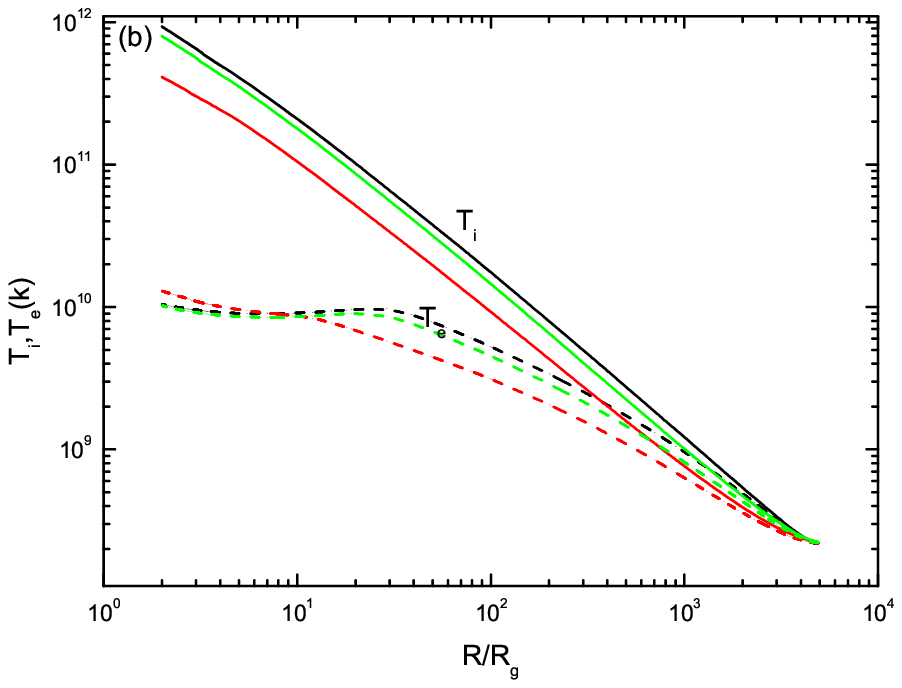}
\includegraphics[width=7.5cm]{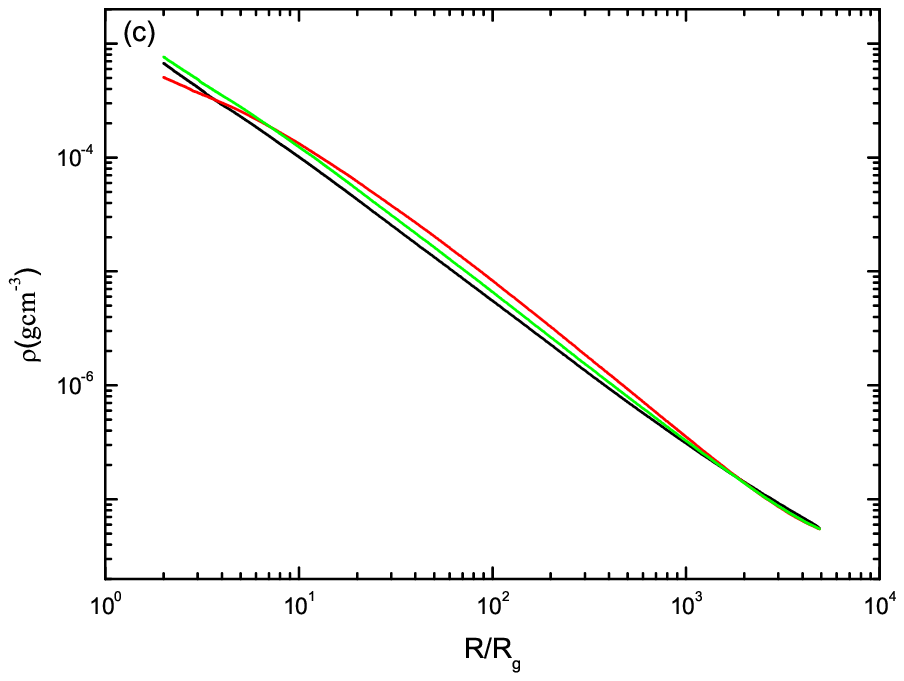}
\includegraphics[width=7.5cm]{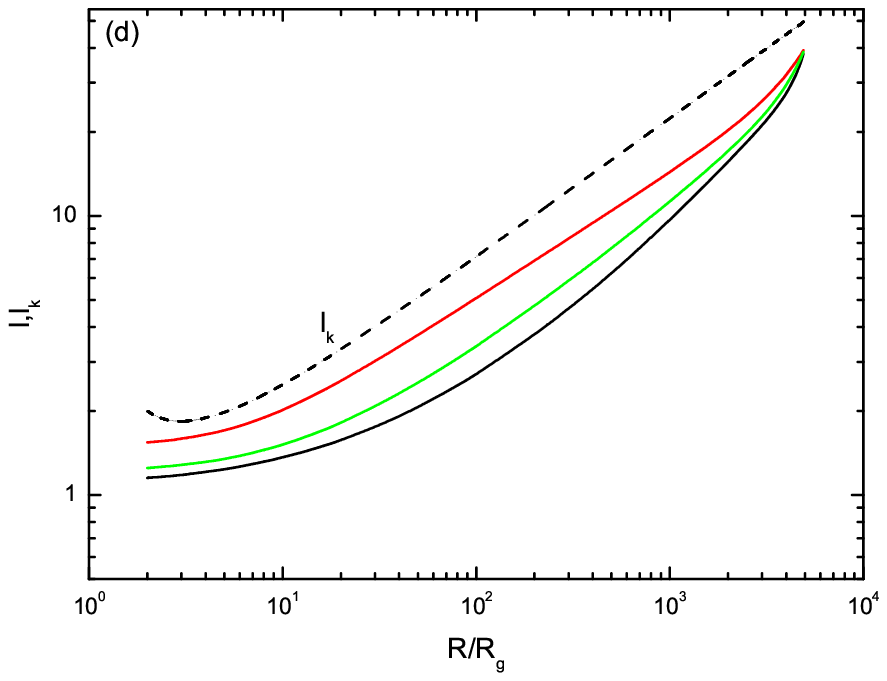}
\caption{The same as Fig. \ref{va1}, but the accretion rate at the
outer radius $\dot{m}_0=10^{-2}$, and $\beta=0.925$, are adopted.
\label{va2}}
\end{figure*}

\begin{figure*}
\includegraphics[width=7.5cm]{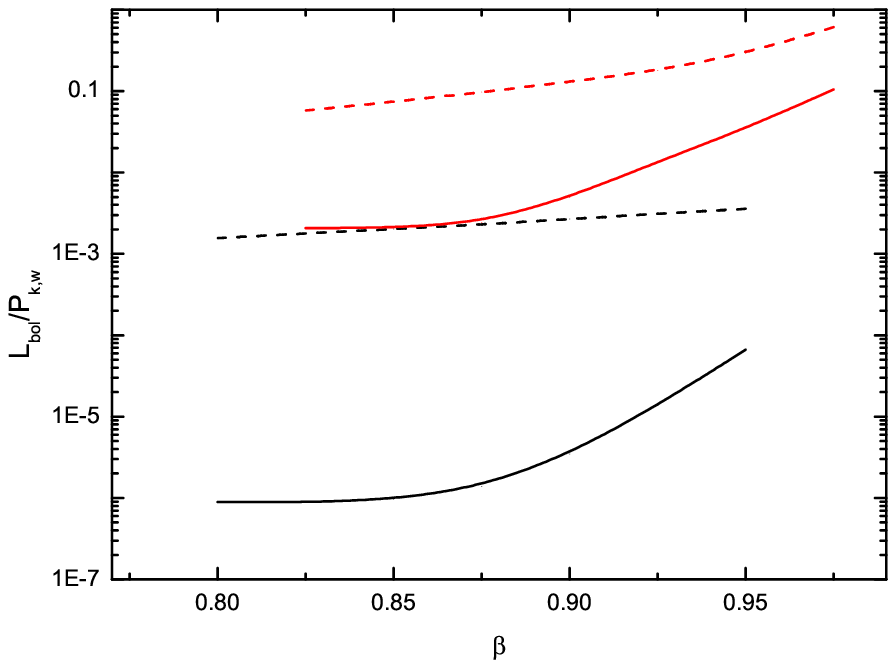}
\caption{The ratio of the bolometric luminosity of accretion disc to
the kinetic power of outflows for different values of $\beta$
($v_{\rm A}=v_{\rm k}$ is adopted). The black lines and red lines
are for the accretion rate at the outer radius $\dot{m}_0=10^{-5}$
and $10^{-2}$ respectively, while the solid lines and dashed lines
correspond to $\zeta=1$ and $4$ respectively. \label{ratio}}
\end{figure*}

We use the Runge-Kutta method to solve a set of five differential
equations (1),(3),(5),(7) and (8) for five variables: $\rho$,
$v_{\rm R}$, $\Omega$, $T_{\rm e}$ and $T_{\rm i}$ with suitable
boundary conditions at the outer radius $R_{\rm out}$. In our
calculations, we adopt the black hole mass $M=10^{8}M_\odot$ for a
typical AGN. The conventional values of the disc parameters:
$\alpha=0.1$ and $\delta=0.1$, are adopted in all our
calculations. The temperature of the ADAF at the outer radius is
adopted as described by the self-similar solution of
\citet{n1995b}. We use a shooting point method in our
calculations. Integrating these five equations from the outer
boundary of the flow at $R=R_{\rm out}$ inwards toward the black
hole, we can obtain the global structure of the accretion flow
passing the sonic point smoothly to the black hole horizon by
tuning the value of radial velocity at $R_{\rm out}$. The outer
radius of the accretion flow $R_{\rm out}=5000R_{\rm g}$ is
adopted. We find that the structure of the ADAF is insensitive to
the outer boundary conditions. As discussed in Sect.
\ref{equations}, the magnetically driven outflow is described by
$v_{\rm A}$, $\zeta$ and $\beta$. The terminal bulk velocity of
the outflow is comparable with the Alf\'{e}n velocity $v_{\rm A}$.
For any unbounded outflows, its bulk velocity should $\ga v_{\rm
K}$, which implies $v_{\rm A}\ga v_{\rm K}$.

The mass accretion rate at the outer radius,
$\dot{M}=10^{-5}\dot{M}_{\rm Edd}$ (the Eddington rate is defined as
$\dot{M}_{\rm Edd}=1.5\times10^{18} M/{{\rm M}_\odot}$ g~s$^{-1}$),
is adopted for the calculations plotted in Figs. \ref{indexs1}--
\ref{va1}. In Fig. \ref{indexs1}, the global solutions for the ADAFs
with outflows are shown with different magnetic field strengths and
distributions (i.e., different values of $\beta$ and $\zeta$), in
which $v_{\rm A}=v_{\rm K}$ is adopted. In the left panel of Fig.
\ref{indexs1}, the radial velocity and the sound speed as functions
of radius with different values of $\beta$ are plotted for
$\zeta=1$. In the right panel of Fig. \ref{indexs1}, the mass
accretion rates as functions of radius are plotted for the global
solutions with different values of $\beta$ and $\zeta$.

In Fig. \ref{zeta1}, we plot different quantities of ADAF
solutions with $\zeta=1$ and 4, respectively, where $v_{\rm
A}=v_{\rm K}$ and $\beta=0.9$ are adopted. As comparison, we also
plot the ADAF solutions without outflows in the same figure. In
the calculations, the terminal velocity of the outflow is a free
parameter. We compare the global ADAF solutions with different
values of $v_{\rm A}$ (i.e., $v_{\rm A}=v_{\rm K}$ and $v_{\rm
A}=2v_{\rm K}$) in Fig. \ref{va1} with $\beta=0.9$ and $\zeta=1$.
In Figs. \ref{indexs2}--\ref{va2}, we plot the results calculated
with a relatively high mass accretion rate, $\dot{m}_0=10^{-2}$,
at the outer radius of the disc.

The ratio of the bolometric luminosity of accretion disc $L_{\rm
bol}$ to kinetic power of outflows $P_{\rm k,w}$ is calculated with
\begin{equation}
{\frac {L_{\rm bol}}{P_{\rm k,w}}}=\int^{R_{\rm out}}_{R_{\rm in}}
q^{-} 4 \pi RH{\rm d}R/\int^{R_{\rm out}}_{R_{\rm in}} (\gamma_{\rm
j}-1)\dot{m}_{\rm w} c^{2} 4\pi R{\rm d}R. \label{ek}
\end{equation}
We plot the ratio $L_{\rm bol}$/$P_{\rm k,w}$ as functions of
magnetic field strength $\beta$ in Fig. \ref{ratio}, where $v_{\rm
A}=v_{\rm K}$ is adopted.

\section{discussion}\label{conclusions and discussion}

A fraction of gases in accretion flow is carried away by the
magnetically driven outflow, which leads to mass accretion rate of
the accretion flow decreasing towards the black hole. In many
previous works, the outflow is induced by assuming the mass
accretion rate $\dot{m}$ to be a power-law dependence of radius
$\dot{m} {\propto} r^s$ \citep*[e.g.,][]{b1999}. In this work, we
obtain global solutions of ADAFs with magnetically driven outflows.
Our results show that the mass accretion rate $\dot{m}$ decreases
towards the black hole, which is close to a power-law $r$-dependence
at larger radii, while it deviates from a power-law $r$-dependence
in the inner region of the ADAF close to the black hole (see Figs.
\ref{indexs1} and \ref{indexs2}). The large-scale magnetic fields
are assumed to thread the accretion flow, which are believed to
accelerate the outflow from the accretion disc. In this work, the
magnetic strength at the disc surface is described by a parameter
$\beta$, which is limited by the gas pressure in the accretion flow.
For comparison, we also plot power-law $r$-dependent accretion rates
in Fig. \ref{indexs1}. Our calculations show that the mass loss rate
in the outflow cannot be very high even if the magnetic fields of
the ADAF are very strong (see Fig. \ref{indexs1}), i.e., the mass
loss rate in the outflow is less than a power-law $r$-dependent
accretion rate with $s\la 0.73$. We also calculate the cases with
high accretion rate, $\dot{m}=10^{-2}$, at the outer radius. It is
found that the results are qualitatively similar to those with
$\dot{m}=10^{-5}$.

For the ADAFs with magnetically driven outflows, the structure of
the accretion flow is significantly altered in the presence of the
outflow. We find that both the ion and electron temperatures of the
ADAF decrease with increasing mass loss rate in the outflow (see
Figs. \ref{zeta1} and \ref{va1}). This is due to the fact that a
fraction of the gravitational energy released in the accretion flow
is tapped to accelerate the outflow, which decreases the heating of
the ADAF. It is easy to understand that the ratio ${L_{\rm
bol}}/{P_{\rm k,w}}$ increases with decreasing magnetic field
strength (see Fig. \ref{ratio}). Based on the magnetically driven
outflow model described in Sect. 2, the kinetic power of the outflow
$P_{\rm k,w}$ can be estimated with
\begin{equation}
P_{\rm k,w} \propto \frac{1}{2} \dot{m}_{\rm w} v_{\rm A}^2
\propto \frac{(R\Omega)^\zeta}{v_{\rm A}^{\zeta-1}}, \label{pk}
\end{equation}
in the non-relativistic limit. It is found that the kinetic power
of the outflow decreases with increasing terminal outflow
velocity, which leads to the structure (e.g., temperatures) of
ADAFs is less altered by the outflows with higher terminal
velocity (see Figs. \ref{va1} and \ref{va2}).

In our present calculations, the magnetic field strength is
estimated with the gas pressure in the ADAF, which is true for the
fields generated with dynamo processes \citep*[see, e.g., the
discussion in][]{l1999}. In this work, the fields are implicitly
assumed to be balanced between diffusion and dynamo/inward
advection in our calculations. The ordered magnetic fields may be
maintained by the drag-in process in the accretion disc, in which
the magnetic field strength is determined by the balance of the
advection and diffusion of the fields in the disc
\citep*[e.g.,][]{1994MNRAS.267..235L}. The magnetic fields can
therefore be stronger than the equipartition limit, however, the
detailed physics is still quite unclear
\citep*[e.g.,][]{1994MNRAS.268.1010L,c2002,2005ApJ...629..960S,2009arXiv0903.3757G}.
In this case, the value $\beta$ (see Eq. \ref{pmag}) describing
the magnetic field strength may vary with radius, which is
determined by the balance between the diffusion and advection of
the fields in the accretion flow
\citep*[e.g.,][]{1994MNRAS.268.1010L}. The mass loss rate in the
outflow can be higher than the results presented in this work, if
the advection of the magnetic fields by the accretion flow is
dominant over the diffusion in the disc.

\section*{acknowledgements}

This work is supported by the NSFC (grants 10773020, 10821302 and
10833002), the CAS (grant KJCX2-YW-T03), and the National Basic
Research Program of China (grant 2009CB824800). S.-L. Li thanks
the support from the Knowledge Innovation Program of Chinese
Academy of Sciences.

\label{lastpage}


\begin{thebibliography}{99}

\bibitem[\protect\citeauthoryear{Armitage}{1998}]{1998ApJ...501L.189A}
Armitage P.~J., 1998, ApJ, 501, L189


\bibitem[\protect\citeauthoryear{Balbus \& Hawley}{1991}]{b1991} Balbus, S., \& Hawley, J. F. 1991, ApJ, 376, 214
\bibitem[\protect\citeauthoryear{Balbus \& Hawley}{1998}]{b1998} Balbus, S., \& Hawley, J. F. 1998, RvMP, 70, 1
\bibitem[\protect\citeauthoryear{Bisnovatyi-Kogan
\& Ruzmaikin}{1976}]{1976Ap&SS..42..401B} Bisnovatyi-Kogan G.~S.,
Ruzmaikin A.~A., 1976, Ap\&SS, 42, 401


\bibitem[\protect\citeauthoryear{Blandford \& Begelman}{1999}]{b1999} Blandford, R. D., \& Begelman, M. C. 1999, MNRAS, 303, L1
\bibitem[\protect\citeauthoryear{Blandford \& Payne}{1982}]{b1982} Blandford, R. D., \& Payne, D. G. 1982, MNRAS, 199, 883
\bibitem[\protect\citeauthoryear{Camenzind}{1986}]{c1986} Camenzind M. 1986, A\&A, 156, 137
\bibitem[\protect\citeauthoryear{Cao}{2002}]{c2002a} Cao, X., 2002, MNRAS, 332, 999
\bibitem[\protect\citeauthoryear{Cao
\& Spruit}{1994}]{1994A&A...287...80C} Cao X., Spruit H.~C., 1994,
A\&A, 287, 80

\bibitem[\protect\citeauthoryear{Cao \& Spruit}{2002}]{c2002} Cao, X., \& Spruit H.C. 2002, A\&A, 385, 289
\bibitem[\protect\citeauthoryear{Gammie et al.}{1999}]{g1999} Gammie, C. F., Narayan, R., \& Blandford, R. D. 1999, ApJ, 516, 177

\bibitem[\protect\citeauthoryear{Guan
\& Gammie}{2009}]{2009arXiv0903.3757G} Guan X., Gammie C.~F., 2009,
arXiv, arXiv:0903.3757


\bibitem[\protect\citeauthoryear{Ho}{2008}]{h2008} Ho, L. C. 2008, ARA\&A, 46, 475
\bibitem[\protect\citeauthoryear{Ichimaru}{1977}]{1977ApJ...214..840I}Ichimaru S., 1977, ApJ, 214, 840
\bibitem[\protect\citeauthoryear{Igumenshchev et al.}{2003}]{i2003} Igumenshchev, I. V., Narayan, R., \& Abramowicz, M. A. 2003, ApJ, 592, 1042
\bibitem[\protect\citeauthoryear{Kato, Kudoh, \& Shibata}{2002}]{2002ApJ...565.1035K} Kato S.~X., Kudoh T., Shibata K., 2002, ApJ, 565, 1035
\bibitem[\protect\citeauthoryear{Koide, Shibata, \& Kudoh}{1999}]{1999ApJ...522..727K} Koide S., Shibata K., Kudoh T., 1999, ApJ, 522, 727

\bibitem[\protect\citeauthoryear{Kudoh \& Shibata}{1995}]{1995ApJ...452L..41K} Kudoh T., Shibata K., 1995, ApJ, 452, L41


\bibitem[\protect\citeauthoryear{Li, Gan, \& Wang}{2009}]{li09} Li
Y., Gan Z.~M., Wang D.~X., New Astron., submitted

\bibitem[\protect\citeauthoryear{Li, Wang,
\& Gan}{2008}]{2008A&A...482....1L} Li Y., Wang D.-X., Gan Z.-M.,
2008, A\&A, 482, 1


\bibitem[\protect\citeauthoryear{Livio et al.}{1999}]{l1999} Livio, M., Ogilvie, G. I., \& Pringle, J. E. 1999, ApJ, 512, 100

\bibitem[\protect\citeauthoryear{Lubow, Papaloizou, \& Pringle}{1994a}]{1994MNRAS.267..235L} Lubow S.~H., Papaloizou
J.~C.~B., Pringle J.~E., 1994, MNRAS, 267, 235
\bibitem[\protect\citeauthoryear{Lubow, Papaloizou, \& Pringle}{1994b}]{1994MNRAS.268.1010L} Lubow S.~H., Papaloizou
J.~C.~B., Pringle J.~E., 1994, MNRAS, 268, 1010


\bibitem[\protect\citeauthoryear{Manmoto}{2000}]{m2000} Manmoto T. 2000, ApJ, 534, 734
\bibitem[\protect\citeauthoryear{McKinney}{2006}]{m2006} McKinney, J. C. 2006, MNRAS, 368, 1561
\bibitem[\protect\citeauthoryear{Michel}{1969}]{1969ApJ...158..727M} Michel
F.~C., 1969, ApJ, 158, 727
\bibitem[\protect\citeauthoryear{McKinney
\& Narayan}{2007a}]{2007MNRAS.375..513M} McKinney J.~C., Narayan
R., 2007a, MNRAS, 375, 513


\bibitem[\protect\citeauthoryear{McKinney
\& Narayan}{2007b}]{2007MNRAS.375..531M} McKinney J.~C., Narayan
R., 2007b, MNRAS, 375, 531



\bibitem[\protect\citeauthoryear{Narayan \& McClintock}{2008}]{2008NewAR..51..733N} Narayan R., McClintock J.~E., 2008, NewAR, 51, 733
\bibitem[\protect\citeauthoryear{Narayan, McKinney,
\& Farmer}{2007}]{2007MNRAS.375..548N} Narayan R., McKinney J.~C.,
Farmer A.~J., 2007, MNRAS, 375, 548



\bibitem[\protect\citeauthoryear{Narayan \& Yi}{1994}]{n1994} Narayan, R., \& Yi, I. 1994, ApJ, 428, L13
\bibitem[\protect\citeauthoryear{Narayan \& Yi}{1995a}]{n1995a} Narayan, R., \& Yi, I. 1995a, ApJ, 444, 231
\bibitem[\protect\citeauthoryear{Narayan \& Yi}{1995b}]{n1995b} Narayan, R., \& Yi, I. 1995b, ApJ, 452, 710





\bibitem[\protect\citeauthoryear{Ogilvie
\& Livio}{1998}]{1998ApJ...499..329O} Ogilvie G.~I., Livio M., 1998,
ApJ, 499, 329
\bibitem[\protect\citeauthoryear{Ogilvie
\& Livio}{2001}]{2001ApJ...553..158O} Ogilvie G.~I., Livio M., 2001,
ApJ, 553, 158

\bibitem[\protect\citeauthoryear{Paczy\'{n}ski \& Wiita}{1980}]{p1980} Paczy\'{n}ski, B., \& Wiita, P. J. 1980, A\&A, 88, 23



\bibitem[\protect\citeauthoryear{Quataert et al.}{1999}]{q1999} Quataert, E., Di Matteo, T., Narayan R., \& Ho, L. C., 1999, ApJ, 525, L89
\bibitem[\protect\citeauthoryear{Quataert \& Narayan}{1999}]{q1999b} Quataert, E., \& Narayan, R. 1999, ApJ, 520, 298
\bibitem[\protect\citeauthoryear{Romanova et
al.}{1998}]{1998ApJ...500..703R} Romanova M.~M., Ustyugova G.~V.,
Koldoba A.~V., Chechetkin V.~M., Lovelace R.~V.~E., 1998, ApJ, 500,
703



\bibitem[\protect\citeauthoryear{Shakura \& Sunyaev}{1973}]{s1973} Shakura, N. I., \& Sunyaev, R. A. 1973, A\&A, 24, 337
\bibitem[\protect\citeauthoryear{Spruit}{1996}]{s1996} Spruit, H. C. 1996, in Physical Processes in Binary Stars, ed. R. A. M. J. Wijers,
M. B. Davis, \& C. A. Tout (Kluwer Dordrecht), 249
\bibitem[\protect\citeauthoryear{Spruit}{2008}]{2008arXiv0804.3096S} Spruit H.~C., 2008, arXiv, arXiv:0804.3096
\bibitem[\protect\citeauthoryear{Spruit
\& Uzdensky}{2005}]{2005ApJ...629..960S} Spruit H.~C., Uzdensky
D.~A., 2005, ApJ, 629, 960

\bibitem[\protect\citeauthoryear{Stepney \& Guilbert}{1983}]{s1983} Stepney, S., \& Guilbert, P. W. 1983, MNRAS,
204, 1269
\bibitem[\protect\citeauthoryear{Stone \& Pringle}{2001}]{s2001} Stone, J. M., \& Pringle, J. E. 2001, MNRAS, 322, 461
\bibitem[\protect\citeauthoryear{Tout \& Pringle}{1996}]{1996MNRAS.281..219T} Tout C.~A., Pringle J.~E.,
1996, MNRAS, 281, 219


\bibitem[\protect\citeauthoryear{Xie \& Yuan}{2008}]{x2008} Xie, F. G., \& Yuan, F. 2008, ApJ, 681, 499
\bibitem[\protect\citeauthoryear{Xue \& Wang}{2005}]{x2005} Xue, L., \& Wang, J. 2005, ApJ, 623, 372
\bibitem[\protect\citeauthoryear{Yuan et al.}{2003}]{y2003} Yuan, F., Quataert, E., \& Narayan, R. 2003, ApJ, 598, 301



\end{thebibliography}
\end{document}